# Effective adhesion strength of specifically bound vesicles

Ana-Sunčana Smith*
E22 Institut für Biophysik, Technische Universität München, D-85748 Garching, Germany

Udo Seifert
II. Institut für Theoretische Physik, Universität Stuttgart, D-70550 Stuttgart, Germany

A theoretical approach has been undertaken in order to model the thermodynamic equilibrium of a vesicle adhering to a flat substrate. The vesicle is treated in a canonical description with a fixed number of sites. A finite number of these sites are occupied by mobile ligands that are capable of interacting with a discrete number of receptors immobilized on the substrate. Explicit consideration of the bending energy of the vesicle shape has shown that the problem of the vesicle shape can be decoupled from the determination of the optimum allocation of ligands over the vesicle. The allocation of bound and free ligands in the vesicle could be determined as a function of the size of the contact zone, the ligand-receptor binding strength and the concentration of the system constituents. Several approximate solutions for different regions of system parameters are determined and in particular, the distinction between receptor-dominated equilibria and ligand-dominated equilibria is found to be important. The crossover between these two types of solutions is found to occur at a critical size of the contact zone. The presented approach enables the calculation of the effective adhesion strength of the vesicle and thus permits meaningful comparisons with relevant experiments as well as connecting the presented model with the proven success of the continuum approach for modeling the shapes of adhering vesicles. The behavior of the effective adhesion strength is analyzed in detail and several approximate expressions for it are given.

## 1. INTRODUCTION

In the past two decades, considerable effort has been invested into understanding the behavior of vesicles consisting of phospholipid bilayers binding to flat substrates. One of the reasons for such development is the fact that vesicles are regularly used as well-defined and controllable model systems for the far more complex process of cell adhesion [1]. Indeed, as cell adhesion plays a central role in key dynamic biological processes such as embryo development, immune response and cancer metastasis, the desire to understand it is hardly surprising.

As adhesion is an important trigger of cell activity [2], nonspecific adhesion controlled by an effective potential between the cell walls must be avoided. Instead, as demonstrated by cell sorting experiments, the highly selective mechanism of cell interactions is based on complementarities between different types of adhesive molecules present on the cell surface [3]. The surface *receptors* are proteins embedded in the lipid bilayer, which constitutes the basic matrix of the outer surface of the cell. These receptors must bind to particular molecular groups of the target cell surface, known as *ligands*, with interaction energies typically in the range of 5 to 20 $k_B$T. Given that the presence of only $10^4$ specific adhesive molecules on the cell surface is sufficient for the normal functioning of the cell, the efficiency of the cell adhesion mechanism is indeed stunning. The receptors and ligands are hidden within a glycocalix, a brush-like macromolecular film rich in carbohydrates that covers the cell surface with a thickness of up to tens of nanometers [3]. The role of this layer is to control precisely the strength of the nonspecific interactions. In the absence of compatible ligands, every embedded receptor contributes to the repelling glycocalix.

Different aspects of bioadhesion have been studied with the help of synthetic models that contain the absolute minimum number of ingredients necessary to mimic cell adhesion [4]. In this respect, various cell models consisting of a vesicle interacting with a substrate have been developed over the last few years [5,6,7]. In all of these models, ligands incorporated into the vesicle membrane are able to bind to receptors immobilized on the substrate. Lipid-coupled polyethyleneglycol molecules (lipopolymers) are used to mimic the cell glycocalix. In all cases, prior to the formation of ligand-receptor bonds, the vesicle settles above the substrate at a distance that is governed by the effective potential acting between the membrane matrix and the substrate. The part of the vesicle membrane that becomes parallel to the substrate in this way is known as the initial contact zone. Due to the fact that, on most occasions, the substrate has been rendered passive by an inert coating and repulsive lipopolymers have been embedded in the vesicle, the part of the membrane in the contact zone is usually only weakly adhered to the substrate and still high enough above the substrate to exhibit strong fluctuations .

The adhesion process associated with the specific biocompatible molecules that follows the initial settling of the vesicle is found to phase segregate the ligand-receptor pairs. Two types of adhesion [8] are observed within the contact zone. The vesicle membrane is either locally trapped in strong adhesion complexes (patches) or it remains in the initial stage of weak adhesion. In the first scenario, the undulations of the membrane are almost totally suppressed whereas, in the remainder of the membrane, fluctuations are still observable. The growth of the patches is believed to be determined by the balance between the osmotic pressures of the lipopolymers that must be expelled from the growing patch and the ligands that contribute to its growth. Empirically, it is only when the formation of patches dominates the overall effective potential that the patches grow beyond the initially established contact zone and induce a first order shape transition [5, 6]. The qualitative explanation of the results observed in these model systems is, however, far from trivial. Indeed, in order to arrive at a satisfactory interpretation of the observed behavior, significant theoretical efforts must be made to understand several aspects of the physics of vesicle adhesion.
An important prerequisite to any theory of vesicle adhesion is naturally an acceptable description of free vesicles. This was provided in the context of a minimal continuum model [9] where the predicted free vesicle shapes were later confirmed by experiment [10]. It was shown in this work that any deformation of a vesicle is constrained by a constant total area and volume. Subsequent advances concerning vesicles bound in a contact potential [11] gave rise to the universal boundary condition which shows that, at the point of contact with the substrate, the vesicle membrane is closing a zero contact angle but with a finite curvature. On the basis of linear extrapolation of the vesicle shape in the vicinity of the substrate, a model connecting the tension in a vesicle with the effective adhesion strength has also been developed [5] and used for the analysis of the contact zone in adhesion experiments [6]. The continuum approach was also later extended to include the influence of gravity [12] to account for the fact that the inner solution of a vesicle is usually associated with a higher mass density than the outer buffer. Finally, in a self-consistent approach, the connection between the thermal fluctuations, the effective tension and the adhesion strength was clarified [13].

It is widely accepted that the continuum models are very successful in explaining the stationary stable shapes of vesicles on the mesoscopic scale. However, by definition, they are unable to account for the discrete nature of the vesicle-substrate interaction, nor have they included the details of the vesicle composition that can be essential for the understanding of the adhesion processes.

Several theoretical models that take the discrete nature of specific binding into account have been developed over time. The thermodynamic considerations of Bell, Dembo and Torney [1, 14, 15] have had a major impact on the understanding of the origins of cell to cell adhesion. Their models were primarily concerned with the balance between the repulsive potentials accorded to the glycocalix, the binding enthalpy and the mixing entropy. Somewhat later, Zuckerman and Bruinsma [16] included membrane-mediated attractions and mapped the statistical model for ligand-receptor interactions to a Coulomb plasma. As a result, they found the ideal mixing state assumed by Bell and co-workers to be unstable against the migration of ligands to the rim of the adhesion plate. They also predicted the enhancement of the membrane adhesion due to fluctuations.

The interplay between lateral phase separation and adhesion of an infinite flat membrane was first considered by Lipowsky and co-workers [17,18]. In the case of vesicles, Komura and Andelman evaluated the mean separation distances between the membrane and the substrate as well as the changes in the height profile, within the contact zone, from patch-like strong adhesion to weak adhesion [19]. Different scenarios for the dynamics of the adhesion process were identified by de Gennes and co-workers and found to depend on the mobility of ligands and receptors as well as on the reaction time associated with binding [20, 21]. Both the statics and the dynamics of colloids adhering specifically to the cell surface have been studied by van Effenterre and Roux though a simple thermodynamic model [22]. Very recently, Coombs et al. have extended previous theoretical approaches [14, 15] to encompass the equilibrium thermodynamics of cell adhesion mediated by two ligand-receptor pairs of different length [23].

In summary, by accounting for the many factors elaborated in the previous section, a good phenomenological coverage of the problem of vesicle adhesion has now been achieved. Unfortunately, however, the results of the theories are often difficult to apply to the actual interpretation of experimental data. The aim of the current paper is therefore to partially rectify this situation. To this end, we attempt to provide a set of self-consistent tools that can be directly applied to measured results. In doing so we hope to start to bridge the gap between experimental and theoretical treatments of vesicle adhesion.

In order to achieve our stated aim, we have simplified the problem of vesicle adhesion to its bare minimum. That is, we consider only the mixing entropy of ligands in the vesicle, the enthalpy of ligand-receptor binding and the bending energy of the entire vesicle shape. The vesicle is treated in a canonical description with a fixed finite number of sites. It will be demonstrated that the finiteness of the system has important repercussions on the behavior of the number of bound ligands and results in two types of equilibria distinguished by the relative concentrations of the ligands and receptors. Furthermore, rather than being restricted to low concentrations of the system constituents, such an approach is suitable for any choice of ligand density in the vesicle and any receptor density on the substrate. We will show that the problem of determining the shape can usually be decoupled from the determination of the number of formed ligand-receptor bonds. We also provide simple approximate analytical solutions, relating the binding energy of a ligand-receptor pair, and their densities in the vesicle and on the substrate respectively, to the number of formed bonds in the contact zone. Particular care has been given to calculations of effective adhesion energy which is often the most important quantity resulting from equivalent experiments.

## 2. THE BASIC ADHESION MODEL

The vesicle surface is initially separated into a region parallel to the substrate (the contact zone) and a region consisting of the remaining part of the vesicle. We allow the interaction between the ligands incorporated into the vesicle and the receptors immobilized on the substrate to take place only within the contact zone. Nevertheless, the ligands in the contact zone are able to exchange with those in the free part of the vesicle. The translational degrees of freedom of the ligands are taken into account by counting the number of microstates (conformations in which the ligands can be distributed over the surface of the vesicle) and finding the most probable macroscopic state. The calculation thus results in the number of bound and free ligands in the two regions of the vesicle at equilibrium.

The following notation is used for the presentation of the model:

- $S_t$ – The total number of sites in the vesicle. One site is of the size of a ligand on the vesicle surface. Within the model, the receptors are of the same size as the ligands.
- $S_c$ – The number of sites forming the contact zone of the vesicle.
- $\rho_r$ – The density of receptors on the substrate. It reflects the extent of surface coverage (e.g. for $\rho_r = 1$ the surface is fully covered, while for $\rho_r = 0$, there are no receptors on the surface).
- $N_t$ – The total number of ligands in a vesicle.
- $N_b$ – The number of ligands that are in the contact zone and bound to receptors.
- $N_f$ – The number of ligands that are in the contact zone and free.
- $N_{free} = N_t - N_b$ – The number of free ligands in the vesicle.

A schematic view of a model vesicle-substrate system is presented in Fig. 1, where the color of a given site reflects whether it contains a ligand, a receptor, or is an empty site. In addition, the bond is formed if, within the contact zone, a site containing a ligand is positioned above a site occupied by a receptor.

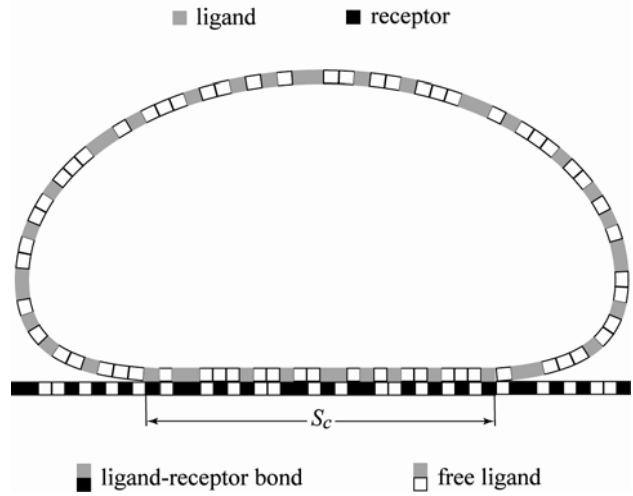

**FIG. 1:** A cross-section of a model system depicting the system constituents. There are a total of $N_t$ grey sites in the vesicle. The contact zone $S_c$ is shown as the part of the vesicle at zero distance from the substrate. The black sites appear with the density $\rho_r$, which results in $\rho_r S_c$ receptors in the contact zone. If, within the contact zone, a grey site is on top of a black site, it contributes to $N_b$ by forming a bond. A grey site over a white site indicates a free ligand in the contact zone that can be associated with $N_f$.

### A. Minimizing the free energy

For a given temperature ($T$), the free energy of the system is given by $F = U - TS$, where $U$ is the internal energy and $S$ is the entropy of the vesicle-substrate system. In the context of the previously outlined model, the internal energy of the system is the sum over all bonds formed in the contact zone:

$$U = -N_b \cdot E_a. \qquad (1)$$

Every formed bond contributes with the adhesion energy $E_a$, which is a measure of the binding strength. The binding strength is a positive quantity and is expressed in units of $k_B T$ ($k_B$ is the Boltzmann constant). The negative sign thus indicates that bond formation is favorable in terms of the total free energy.

The entropy is calculated by counting all possible conformations ($\Omega$) of the positions of the ligands in the vesicle. Under the previously described assumptions, $\Omega$ is the number of combinations in which one can place: $N_t - N_b - N_f$ ligands in the free part of the vesicle on $S_t - S_c$ positions, $N_f$ free ligands on $S_c(1-\rho_r)$ sites not occupied by receptors, and $N_b$ bound ligands on $\rho_r \cdot S_c$ receptor sites:

$$\Omega = \binom{S_t - S_c}{N_t - N_b - N_f}\binom{S_c(1-\rho_r)}{N_f}\binom{\rho_r \cdot S_c}{N_b}. \qquad (2)$$

The entropy is given by $S = k_B T \ln \Omega$ and can be calculated with the use of the Stirling formula for factorials of large numbers. The resulting expression is then analogous to the standard mixing entropy term [24], but due to its length will not be presented here.

Minimizing the free energy, $F = U - TS$ with respect to $N_b$ and $N_f$, should provide the ideal fraction of bound and free ligand molecules. The first important relation resulting from the minimization can be cast in the form of the "density equation":

$$\frac{N_t - N_f - N_b}{S_t - S_c} = \frac{N_f}{S_c(1-\rho_r)} \qquad (3)$$

The left side of this equation is the density of ligands in the free part of the vesicle. Similarly, the right hand side is the density of free ligands in the adhesion zone over sites not occupied by receptors. As the free energy of a ligand in the upper part of the vesicle is, within the current model, the same as the free energy of an unbound ligand in the contact zone, a violation of the density equation (3) would lead to unequal lateral pressures of free ligands inside and outside the contact zone, and hence, to the loss of a stationary solution.

The second important relation resulting from the minimization shows the influence of the binding strength on the allocation of bound and free ligands within the contact zone:

$$\frac{(1-\rho_r)\cdot S_c}{N_f} - 1 = e^{E_a}\left(\frac{\rho_r \cdot S_c}{N_b} - 1\right) \qquad (4)$$

On this occasion, it is the densities of the free and bound ligands within the contact zone, (weighed by the Boltzmann factor of each state) which are being equilibrated. Solving Eq. (3) and Eq. (4) simultaneously for $N_f$ and $N_b$ results in the optimum allocation of ligands in the vesicle, for a given size of the contact zone.

Further inspection of the free energy reveals that the density equation (3) can be obtained directly by minimizing the free energy with respect to the size of the contact zone. As a result, although the free energy depends on three variables, one of the equations emerging from the minimization of the free energy with respect to these three variables is linearly dependent on the remaining two. Moreover, it is easy to show that the free energy is a decreasing function of $S_c$, leading to a boundary minimum with respect to the same variable. In order to reach the thermodynamic equilibrium, the vesicle will thus maximize its area of contact with the substrate. However, the size of the contact zone is restricted by the volume and area constraints and the bending energy of the vesicle. Furthermore, the ratio between the bending energy and the free energy calculated herein scales inversely with the total number of sites in the vesicle ($S_t$), which is usually a very large number. Hence, the magnitude of the bending energy term is very small in comparison with the internal energy and the entropy, and can generally be omitted from the calculation. The only exception is when the shape of the vesicle approaches the shape of a spherical cap. Due to the constraints on the total volume and area of the vesicle, it is this shape that limits the size of the contact zone [11]. For a spherical cap, a well-defined contact angle is formed with the substrate. This causes the bending energy to diverge and induces a stable boundary minimum in the total free energy of the vesicle adhering to the substrate. Thus in the thermodynamic equilibrium, the vesicle shape is always that of the spherical cap, with an optimum number of bound and free ligands in the contact zone. Depending on the coverage density and the binding affinity of the ligand-receptor pair, the density of bonds can vary from very low to very high.

If, for some reason, the adhesion process is very slow, the free energy will relax with respect to the number of ligands much faster than with respect to the size of the contact zone. Hence, a constrained equilibrium, where the vesicle shape is not that of a spherical cap, can be experimentally observed. In that case, the shape of the vesicle can in principle be determined with the use of the continuous models [11], where the free energy must be minimized with respect to the given size of the contact zone. The number of bound ligands corresponding to this size of the contact zone can be determined as described in following sections.

**B. The number of ligands in the contact zone**

In order to determine the allocation of ligands over the vesicle, the system, consisting of Eqs. (3) and (4) can be solved analytically, which results in two sets of solutions for $N_b$ and $N_f$. However, there is only one physically relevant set from which we present the resulting allocation function for the number of bound ligands:

$$N_b = \frac{N_t + \rho_r S_c}{2} + \frac{S_t}{2 \cdot (e^{Ea} - 1)}$$
$$- \frac{\sqrt{e^{2Ea}(N_t - \rho_r S_c)^2 - 2e^{Ea}[N_t^2 + \rho_r^2 S_c^2 - (N_t + \rho_r S_c)S_t] + (N_t + \rho_r S_c - S_t)^2}}{2 \cdot (e^{Ea} - 1)} \quad . \quad (5)$$

The number of free ligands within the contact zone ($N_f$) can be easily obtained by substitution of Eq. (5) into the density equation (3). The total number of free ligands in the vesicle is simply $N_{free} = N_t - N_b$.

*(a) Limiting behaviors.* If there is no interaction between the ligand and the receptor, *i.e.* $E_a = 0$, the ligands are uniformly allocated over the vesicle. Hence the number of bound and free ligands in the contact zone is scaled by the number of sites containing receptors and the number of empty sites, respectively (first column in Table I). As the binding strength increases, all allocation functions exhibit considerable deviations from their values at $E_a = 0$. However, the allocations reach their saturation values, characteristic for the limit of $E_a \to \infty$ (see columns 2 and 3 in Table I) surprisingly quickly, typically at binding strengths between 10 and 15 $k_B T$. For large binding strengths, the allocation functions are limited by the concentrations of the vesicle and substrate constituents. If there are more ligands available than receptors in the contact zone ($N_t > \rho_r S_c$), all receptors will be bound (see the second column in Table I). If, on the other hand, the total number of receptors in the contact zone is larger than the total number of ligands in the vesicle ($\rho_r S_c > N_t$), all of the ligands will be bound (see the last column in Table I). Importantly, which of the two limits is applicable to a certain composition of the vesicle and the substrate will depend on the chosen size of the contact zone $S_c$. Actually, there is a critical size of the contact zone $S_c^*$ for which the number of ligands in the vesicle is same as the number of receptors in the contact zone:

$$S_c^* = \frac{N_t}{\rho_r}$$

It is for this size of the contact zone that the two limits at $E_a \to \infty$ become equivalent.

**TABLE I.** The limits of the allocation functions.

|  | $E_a = 0$ | $E_a \to \infty$ | |
|---|---|---|---|
|  |  | $S_c < S_c^*$ | $S_c > S_c^*$ |
| $N_b$ | $\dfrac{\rho_r S_c N_t}{S_t}$ | $\rho_r S_c$ | $N_t$ |
| $N_f$ | $\dfrac{(1-\rho_r) S_c N_t}{S_t}$ | $\dfrac{S_c (1-\rho_r)(N_t - \rho_r S_c)}{(S_t - \rho_r S_c)}$ | 0 |
| $N_{free}$ | $\dfrac{(S_t - \rho_r S_c) N_t}{S_t}$ | $N_t - \rho_r S_c$ | 0 |

*(b) Overall behavior.* The consequences of the existence of two limits for the number of ligands inside and outside the contact zone when $E_a \to \infty$, strongly affects the behavior of the allocation functions. In particular, if there are more ligands in the system than receptors ($N_t > \rho_r S_c$), the balance of the system entropy and enthalpy will be dominated by the lack of receptors for any given binding strength $E_a$. This situation will be referred to as a ***receptor-dominated equilibrium***. On the other hand, the presence of more receptors than ligands leads to a stable solution that is limited by the total number of ligands on the vesicle surface and will result in a so-called ***ligand-dominated equilibrium***.

The transition from one class of equilibria to the other can be obviously achieved by changing the osmotic conditions, the protein contributions in the system or by inducing a detachment process. However, if $N_t$ and $\rho_r$ are chosen (resulting in a particular $S_c^*$), and $S_c$ is fixed, the only remaining parameter is the ligand-receptor binding strength $E_a$. Specifically, if $S_c < S_c^*$, and one plots the allocation functions against $E_a$, each curve will correspond to a set of a receptor-dominated equilibria (left panel in Fig. 2). In the case of $S_c > S_c^*$ the system adopts into one of the ligand-dominated equilibria (right panel in Fig. 2).

It should also be expected that the equilibrium of the system will be influenced considerably by the strength of the ligand-receptor binding. Inspection of Fig. 2 shows that at small binding strengths, the number of bound ligands in the vesicle is smaller than the number of free ones ($N_b < N_{free}$). As the strength increases, the majority of ligands will become bound to the receptors. This is a direct consequence of the fact that the total number of ligands in the vesicle is finite. The characteristic binding strength at which the number of bound ligands begins to exceed that of free ligands in the vesicle can be found by solving $N_{free} = N_b = N_t / 2$ which results in:

$$E_a = \ln \frac{2S_t - N_t - 2\rho_r S_c}{2\rho_r S_c - N_t}. \tag{6}$$

The crossing will occur as long as $\rho_r S_c > N_t/2$. At $\rho_r S_c < N_t/2$ the expression for number of bound ligands saturates at a value below the limit for the total number of free ligands in the vesicle.

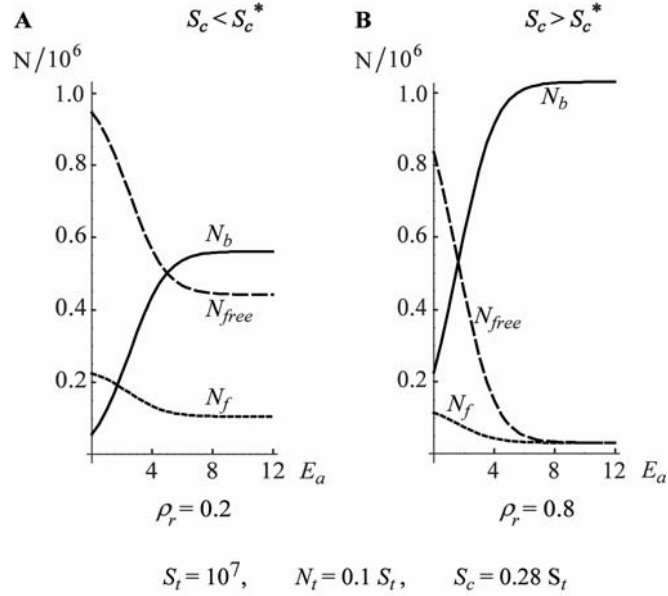

**FIG. 2:** The allocation in the contact zone of bound (full line) and free ligands (short dashed line) are presented together with the total number of free ligands in the vesicle (long dashed line) as a function of the binding strength $E_a$. In both panels, the case of a vesicle that has 85% of the volume of a sphere with the same surface area (reduced volume $v = 0.85$) is considered. In thermodynamic equilibrium, this leads to a spherical cap with a contact zone that comprises 28% of the total vesicle surface area ($S_c = 0.28\, S_t$). Furthermore, if the diameter of the vesicle is 10 $\mu$m and the ligand incorporated has the gyration radius of 3.5 nm (as does the commonly used sialyl-Lewis$^X$-glycosphingolipid) then the total number of sites $S_t$ in the vesicle is about $10^8$. If the vesicles are prepared with the number concentration of about $10^{-3}$ of the ligand with respect to the lipid, 10% of the vesicle surface will be covered by ligands ($N_t = 0.1\, S_t$). The coverage of the substrate by receptors is chosen to be (A) 20% ($\rho_r = 0.2$) so a set of receptor-dominated equilibria is obtained and (B) 80% ($\rho_r = 0.8$) which results in a set of ligand-dominated equilibria. The coverage is indicated directly below the graphs whereas the other parameters can be found in the bottom line of the figure.

For coverages of the substrate less than $\rho_r = 0.5$, the allocation function for bound molecules ($N_b$) intersects that for free receptors in the contact zone ($N_f$). By solving $N_f = N_b$, the following crossover binding strength is obtained:

$$E_a = \ln \frac{S_c(1-\rho_r)(S_t - N_t + (1-2\rho_r)S_c)}{\rho_r S_c(S_t + (1-2\rho_r)S_c) - N_t S_c(1-\rho_r)}. \tag{7}$$

This crossover vanishes at $\rho_r = 0.5$ when, due to the equipartition of ligands imposed by the density equation (3), $N_f = N_b$ at $E_a = 0$. Furthermore, if the coverage is very low, the expression for the number of bound ligands saturates below that for the number of free ligands in the contact zone and the crossing does not occur for any value of $E_a$. In this case, the concentration of free ligands in the contact zone is much larger than the concentration of bound ones.

Some transitionary behavior of the allocation functions (and of other quantities derived by the use of allocation functions) should be expected for $S_c = S_c^*$. Due to the geometrical constraint on the maximal size of the contact zone for a given total area of the vesicle, such transitions should disappear when $S_c^* > 0.5 S_t$ as the system always relaxes into a receptor-dominated equilibrium. The analysis of such a transition, the behavioral regimes of the allocation functions and the development of several useful approximate relations is given in the Appendix A.

### 3. THE EFFECTIVE ADHESION STRENGTH

Vesicle adhesion is often regarded as a wetting phenomenon where the spreading pressure of a vesicle is determined as the work of the system to induce changes in the contact area with the substrate. Several experimental studies have reported measurements of the adhesion strength of specifically adhered vesicles based on the usage of the Young's law [5, 6]. Moreover, previous theoretical investigations have shown that the shape of the vesicle can be understood by adhesion in a continuous (contact) potential [11]. In these models, the average adhesion strength is constant, independent of $S_c$ or given externally. To make a link between the present model and the previously used approaches, it is necessary to calculate the effective adhesive potential ($W$) resulting from numerous local bindings. Within the canonical approach undertaken herein, the average adhesion strength becomes a function of the size of the contact zone, with a functional dependence that can be determined by calculating:

$$W \equiv -\frac{1}{a}\frac{\partial F}{\partial S_c}\bigg|_{\rho_r} \equiv \frac{\omega}{a}. \qquad (8)$$

Here $a$ is the area of the unit cell determined by the size of a ligand. For a chosen value of $S_c$, it is found that:

$$\omega = \rho_r \cdot \left( \ln\frac{(1-\rho_r)S_c - N_f}{(1-\rho_r)S_c} - \ln\frac{\rho_r S_c - N_b}{\rho_r \cdot S_c} \right) \qquad (9)$$

To obtain $\omega$ for a given size of the contact zone, $N_b$ and $N_f$ must be calculated from Eqs. (3-5). The first term on the right hand side of Eq. (9) is the natural logarithm of the density of ligand-free sites in the part of the contact zone unoccupied by receptors, whereas the second term is the logarithm of the density of ligand-free sites in the part of the contact zone occupied by receptors. As the chemical potential is the logarithm of a density, $\omega$ is the result of an imbalance between two chemical potentials of *empty* sites, weighed by the density of receptors.

The dependence of the effective adhesion strength ($W$) on the ligand size ($a$) results in $\omega$ expressed in units of $k_b T$. For a ligand with a gyration radius of 3.5 nm, $\omega = 1$ leads to $W$ of the order of $10^{-5}$ N/m. Moreover, the conversion to adhesion strengths $w$, as employed in continuous models [11-13,26,27], can be achieved by the relation $w = \omega \cdot 4\pi\kappa / S_t$. For a standard bending rigidity of the membrane $\kappa = 100$ $k_B T$ and $S_t = 10^7$, this conversion would lead to $w \approx 10^{-4} \omega$.

*(a) Limiting behavior and the dependence on the binding strength of the ligand-receptor pair.* Several important properties of $\omega$ are inherent from the allocation functions and are a result of the existence of the two sorts of equilibria (dominated by either by ligands or receptors) and their limits when $E_a \to \infty$. However, for both types of equilibria, $\omega$ is vanishing as $E_a \rho_r N_t / S_t$ as $E_a \to 0$. In the receptor-dominated equilibria ($S_c < S_c^*$), a diverging linear regime (presented in the left panel of Fig. 3) is found to dominate the behavior of $\omega$:

$$\omega \approx \rho_r E_a + \rho_r \ln \frac{N_t}{2(S_t - N_t)}, \text{ for } E_a \to \infty. \quad (10)$$

In the ligand-dominated equilibria ($S_c > S_c^*$) inspection of the right panel in Fig. 3 reveals a rapid increase of $\omega$ followed by convergence to a finite value:

$$\omega \approx \rho_r \ln \frac{\rho_r S_c}{\rho_r S_c - N_t}, \text{ for } E_a \to \infty. \quad (11)$$

After a certain binding strength, all of the ligands in the vesicle become bound. Hence, in this regime, further increasing the ligand-receptor binding strength does not influence the average adhesion strength.

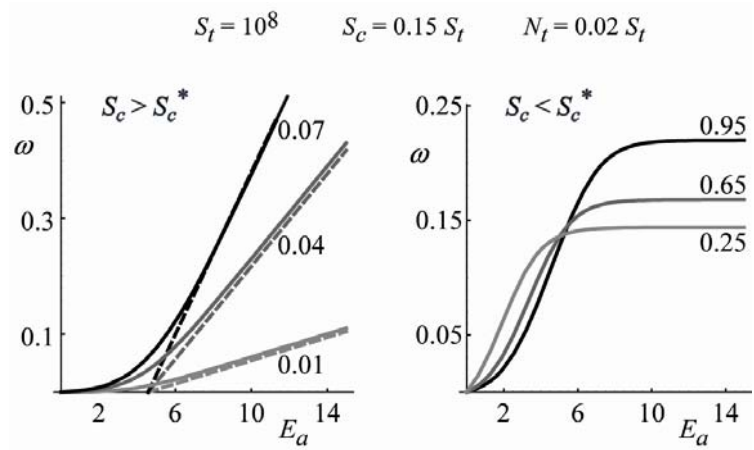

**FIG. 3:** Effective adhesion strength as a function of the ligand-receptor binding strength for different coverage densities (numbers indicated next to curves). Left: Diverging of the effective adhesion strength characteristic for the receptor-dominated equilibria. The full solution (lines) and the asymptote from Eq. (10) (dashed lines) are shown. Right: The saturation of the average adhesions strength is characteristic for the ligand-dominated equilibria. The effective adhesion strength for this type of equilibrium converges to values given by Eq. (11).

*(b) Dependence on the size of the adhesion plate.* Detaching the vesicle by means of local force application, flow, or the insertion of antibodies is usually associated with changes in the size of the contact zone. Hence, the work on the vesicle performed by any of these means can be evaluated from the change in the effective adhesion strength. Thus, it is particularly important to understand the dependence of the adhesion strength on the size of the contact zone presented in Fig. 4.

It is easy to show that $\omega$ is a monotonically decreasing function of $S_c$ independently of the choice of other parameters. This is a consequence of the fact that the density of bound ligands never increases in response to an increase in the size of the contact zone (at a given coverage, binding strength and total number of ligands in the vesicle). Instead, the large changes in $\omega$ are a consequence of the changes of the derivative of the density with respect to $S_c$. When the allocation function of bound ligands reaches its limiting values, imposed by the composition of the vesicle and the substrate, the density of bonds is independent of the size of the contact zone. The maximum value in the effective adhesion strength ($\omega_0$) is

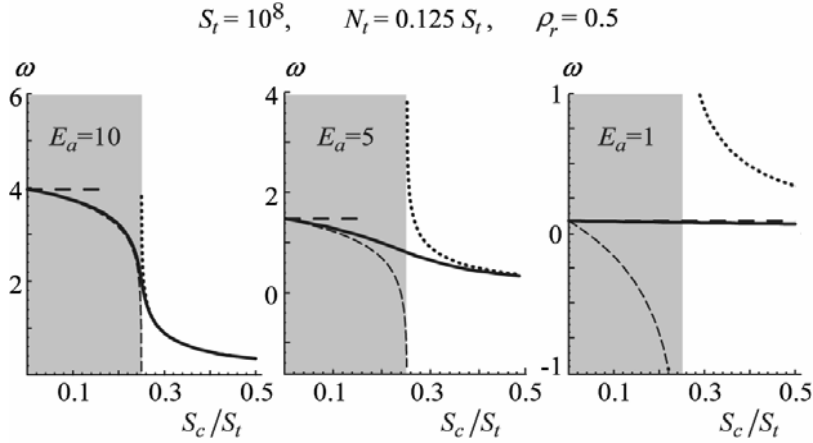

**FIG. 4:** Effective adhesion strength as a function of the size of the contact zone. Due to geometrical constraints, the size of the contact zone is restricted to half of the total vesicle area. The region shaded in gray is depicting the region characteristic for receptor-dominated equilibria, whereas the region of parameters resulting in ligand-dominated equilibria is shown with a white background. Left: Nonlinear behavior with a crossover at $S_c^*$ described with approximate solutions from Eq. (10) (dotted lines) and Eq. (13) (short dashed lines) characterize the $\omega$ for $E_a$ Middle: For intermediate binding strengths of the ligand-receptor pair, agreements with the approximate solutions is obtained only in some parts of the curves. These parts are imposed by the coverage density of the substrate. Right: For low binding strengths, the approximate solutions are not valid. The effective adhesion strength is a slowly varying function and its magnitude is given by Eq. (12), along the whole range of sizes of the contact zone.

reached at $S_c=0$ and is calculated to be:

$$\omega_0 = \rho_r \ln\left[\left(e^{E_a}-1\right)\frac{N_t}{S_t}+1\right]. \tag{12}$$

Interestingly, if Eq. (12) is plotted as a function of $E_a$, then excellent overlap is obtained with the curves presented in Fig. 3 for small $E_a$ for both types of equilibria (data not presented). Furthermore, for large $E_a$ and in the receptor-dominated equilibria (left panel in Fig. 3), $\omega_0$ increases linearly with the same slope as predicted by the asymptote Eq. (10), but with a somewhat underestimated offset of $\omega_0$. However, for a choice of parameters resulting in this receptor-dominated equilibria, at low coverage $\omega_0$ provides a very good approximation to the full solution of $\omega$ over the entire range of $E_a$.

It is important to notice that, when the adhesion strength is presented as a function of the size of the contact zone (Fig. 4), an inflection point occurs at $S_c^* = N_t / \rho_r$. In the region resulting in ligand-dominated equilibria, Eq. (11) (shown with a dotted line in Fig. 4) can be used to approximate the real solution for $\omega$. An analogous approximate relation is determined in the region producing receptor-dominated equilibria where $S_c < S_c^*$:

$$\omega \cong \omega_0 - \rho_r \ln \frac{N_t}{N_t - \rho_r S_c}. \tag{13}$$

The comparison of Eq. (13) (short dotted lines) with the real solution is given in Fig. 4. For large $E_a$, both Eq.(13) and Eq. (11) are found to be very useful, as shown on the left panel in Fig. 4. In the intermediate range of binding strengths (middle panel of Fig. 4), the region around the inflection point is badly reproduced by the approximate solutions. The width of this region depends entirely on the binding strength $E_a$. However, the position of the inflection point can be regulated by changing the density of ligands in the vesicle and the receptors on the surface. Both an increase of $N_t$ and a decrease of $\rho_r$ are capable of translating $S_c^*$ to larger values. Therefore, for such intermediate values of $E_a$, Eq. (13) provides a good approximation for small sizes of the contact zone. Conversely, a decrease of $N_t$ or an increase of $\rho_r$ shifts $S_c^*$ to smaller values. This permits the use of Eq. (10) at large sizes of the contact zone.

For weak ligand-receptor pairs, the density of ligands in the contact zone is almost constant for any size of the contact zone and $\omega$ experiences only very small changes. Though neither Eq. (10) nor Eq. (13) is applicable in this range of $E_a$, the effective adhesion strength can be approximated with $\omega_0$ as shown in the right panel of Fig. 4.

## 4. DISCUSSION AND SUMMARY

The aim of this work was to produce a tool simple enough to be manageable and applicable to experiments while at the same time retaining sufficient sophistication to account for the most important contributions to the free energy in the adhesion process of vesicles. Hence, we have calculated the thermodynamic equilibrium for a vesicle containing ligands capable of specific binding to receptors on the substrate. Although this work is, in spirit, based on the same physical ideas as those employed in the well known studies of Dembo, Bell and Torney [ , ], we have succeeded to identify and characterize, in an unencumbered manner, some important regimes in vesicle adhesion driven by specific binders.

The results of the calculations show that the choice of the statistical ensemble is an important issue. In particular, the experimental reality is such that the contact zone of the vesicle with the substrate is usually relatively large, so that the final number of receptors in the contact zone is at least comparable to the total number of ligands in the vesicle [6, 25]. The large adhesion patches (of the order of μm$^2$) obtained in these experiments, indicate the formation of numerous ligand-receptor bonds. However, the ligands participating in these bonds have had to diffuse from the free part of the vesicle (hence the observation of diffusion limited adhesion). As the experiments are performed with a constant total number of ligands in the vesicle, the concentration of ligands in the upper part of the vesicle has therefore had to undergo considerable reduction upon adhesion. Under these circumstances, treating the adhesion in a grand canonical model where the ligands are coupled to a bath of constant chemical potential is not correct. Rather, a canonical statistical ensemble should be imposed on the vesicles. Regardless of the parameterization of the canonical ensemble, the condition of thermal equilibrium dictates that the chemical potential of the ligands in the contact zone and in the free part of the vesicle are equilibrated. The entropic cost for depletion in one region is balanced by the gain in the internal energy in the other region. Even if a given molecule is in the region of increased density, this molecule has on average, no incentive no penetrate the depleted region. Hence, differences in densities between the two regions cannot be directly interpreted as a lateral osmotic pressure. Such pressure would arise from unbalanced chemical potentials and would be contrary to the equilibrium condition.

A direct consequence of the canonical treatment of vesicle adhesion is the identification of two types of equilibria dominated by the contribution of ligands and receptors, respectively. When the number of receptors in the applicable contact zone is same the total number of ligands in the vesicle, the system undergoes a crossover between the two types of equilibria. The existence of these two types of equilibria is a result of the finite reservoirs of ligands and receptors in the system and has implication not only on the allocation of bound and free ligands in the vesicle but also on the behavior of the effective adhesion strength.

Our calculations suggest that the use of ligand-receptor pairs associated with a strong binding constant (as in the case of the biotin-streptavidin or RGD-integrin pairs), would lead the vesicle-substrate system into an equilibrium described by the one of two limits of the allocation function of bound ligands for $E_a \to \infty$. Hence, there should basically be either no free ligands in the vesicle or no free receptors on the substrate.

The equilibrium of the system can be found either by finding the optimum number allocations (as in the case in the presented model) or by equilibration of chemical potentials. The result is independent of the procedure. However, in a possible experiment where one of the system parameters can be continuously tuned, the pathway will depend on whether each new state is associated with constant chemical potential or a constant number of particles. Changing the size of the contact zone by means of adjusting the osmotic conditions of the buffer solution, or influencing the binding strength by changing the content of the buffer are two possible ways that could be used for exploring the equilibrium adhesion of a single vesicle. As such processes are associated with a constant number of ligands in the vesicle, the number allocation functions should be employed to interpret the measured changes. Manipulating parameters of the system such as $S_c$ or $E_a$ would, according to the presented calculations, be equivalent to moving along one of the lines presented in Fig. 5 or Fig. 6 in the Appendix A, respectively.

An important fact emerging from the model is that the bending energy can be virtually omitted from the calculations. The bending energy is a function of the membrane elastic modulus $\kappa$ and seldomly exceeds $10^{-5}$ $k_BT$ per site on the vesicle. On the other hand, the contribution of each ligand in a vesicle is of the order of 1 $k_BT$. As the number of ligands is very large, it is clear that the bending is not of comparable magnitude. Hence, as long as the shape of the vesicle is not that of the spherical cap, deformations of the membrane are energetically inexpensive, provided that the ratio between the surface area of the contact zone and the free part of the vesicle remains unchanged. This explains the stability of strongly deformed membranes balanced by an agglomeration of ligand-receptor bonds at the edge of the contact zone, as often observed in experiments on weakly adhered vesicles. As the strength of adhesion (or the density of bonds) increases and the spherical cap is approached, the tension in the vesicle becomes large and unusual deformations become energetically expensive and unobservable. Indeed, in the experiments where the shape of a vesicle is a spherical cap, the contact zone is observed to be discoid with no pronounced deformations whatsoever [6].

Due to the divergence of the bending energy, we were able to identify a boundary minimum with respect to the size of the contact zone. It leads to a thermodynamic equilibrium in which the contact zone is maximized and the vesicle always assumes the shape of a spherical cap, as seen in some experiments [6]. Depending on the coverage of the substrate and the density of ligands, the number of bound molecules can be determined by Eq. 5 or one of its approximate solutions $N_b^{SIG}$ or $N_b^{LIN}$ (see Appendix A).

Despite the predicted existence of the boundary minimum, there are quite some experimental situations, where the vesicle appears to be in its equilibrium state without assuming the shape of the spherical cap [5]. The adhesion process associated with such a state is usually slow and stepwise, and should be expected when the probability for bond formation is reduced, due to a low coverage or a low fraction of ligands in the vesicle. Technically, the slow equilibration leads to a relaxation of the free energy with respect to the number of ligands in the contact zone but not with respect to the size of the contact zone. In this constrained equilibrium, the allocation functions resulting from the minimization are still valid, but the size of the contact zone is not determined by the bending divergence but by factors such as the non-specific interaction potential, shape fluctuations and the probability for bond formation. The shape of the vesicle in this constrained equilibrium can be determined by the use of a continuum model [11] where the bending energy must be minimized for a chosen size of the adhesion zone. However, it is important to emphasize that the proposed allocations can be applied to both the thermodynamic and the constrained equilibria.

The densities of bound and free ligands in the contact zone are found to be responsible for the strength of the effective adhesive potential. This potential, in thermodynamic equilibrium, is equivalent to the spreading pressure obtained from the Young-Dupré law for liquid droplets. However, in the constrained equilibrium, the contact angle of the vesicle with the substrate is not well defined and so the Young-Dupré law is not valid. Under these circumstances, the effective adhesion strength and the shape of the vesicle can be calculated by minimizing the bending energy of a vesicle for a given size of the contact zone by the methods employed in continuous models. The effective adhesion strength of constrained equilibria is dominated by non-specific interactions. Hence, the $\omega$ originating from the current model should be interpreted as the bond contribution to the total effective adhesion strength.

The calculation of the effective adhesion strength provides a bridge between models based on a continuous potential and those based on discrete specific binding. As shown in [26], pulling on vesicles in constrained equilibria will result in continuous shape deformations, whereas pulling on vesicles in thermodynamic equilibrium results in tether formation [27]. Thus, the knowledge of the effective potential enables the determination of the vesicle shape (from continuous models) and the number of bonds in the contact zone (from the current discrete model) [28]. In addition, the model outlined herein can be used for the identification of the main mechanisms in which a competitive binder (an antibody), injected into the surrounding buffer, acts on specifically adhered vesicles [29]. In conclusion, although the presented model is relatively simple, it produces practical results that are widely applicable to the interpretation of experimental data.

## ACKNOWLEDGMENTS

We would like to thank E. Sackmann for motivating this work and numerous helpful discussions. A.-S. S. would like to thank the Hochschul- und Wissenschaftsprogramm (HWP II) for the financial assistance, as well as the Sonderforschungsbereich 465/C4 for support.

## APPENDIX A

In this Appendix we present different aspects of Eq. (5) in terms of their dependence on the size of the contact zone and the binding strength of the ligand receptor pair. The results of this discussion will be summarized in a table in the end of thie Appendix.

### *1. Dependence of the allocation functions on the size of the adhesion plate*

The size of the contact zone can vary from zero (for spherical vesicles) to a maximum one half of the total area (completely deflated vesicle). Due to this geometrical constraint on $S_c$ and the two classes of equilibria for vesicle adhesion, two distinct types of behavior of the allocation function for the number of bound ligands can be identified in Fig. 5, (left and the right panel, respectively).

Several combinations of system parameters result in linear regimes of $N_b$ that can be approximated by the following expression:

$$N_b^{LIN} = \frac{(e^{E_a}+1) \cdot N_t}{(e^{E_a}-1) \cdot N_t + 2S_t} \cdot \rho_r S_c. \tag{A1}$$

Equation (A1) is obtained by connecting the zero point ( $N_b = S_c = 0$ ) with the point at which $N_b = 0.5 \cdot N_t$ and is presented in Fig. 5 with dotted lines.

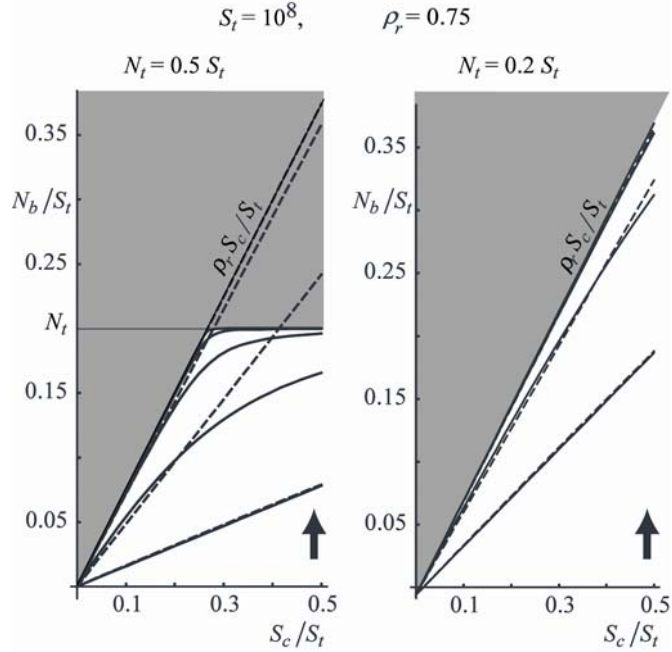

**FIG. 5:** The number of bound ligands as a function of the size of the adhesion plate for a set of binding strengths ( $E_a = 0.1,\ 2.5,\ 5.0,\ 7.5,\ 10.0$ ), for constant $S_t$ and $\rho_r$. The arrows indicate the direction of the increase of the binding strength. The total number of ligands is indicated above the left and right panels, respectively. The regions in the graph where $N_b$ has solutions are depicted with white background, whereas the remaining part of the parameter space is shown in gray. The approximate $N_b^{LIN}$ solution (short dotted lines) is presented together with the real solution (full lines).

Linearizing $N_b$ can also be performed by using the slopes at $S_c = 0$:

$$N_b^0 \equiv \left.\frac{\partial N_b}{\partial S_c}\right|_{S_c=0} \cdot S_c = \frac{e^{E_a} N_t}{S_t - (e^{E_a} - 1)N_t} \rho_r S_c. \tag{A2}$$

Although $N_b^0$ is a valid approximation of Eq. (5), the approximation given by Eq. (A1) is found to be the simplest expression that accurately describes the real solution from Eq. (5) for the widest choice of system parameters.

*(a) $S_c^* < 0.5 S_t$ regime.* This regime is presented in the left panel of Fig. 5 where the boundaries of the region where solutions for $N_b$ can be found are the horizontal $N_t$ line and the $\rho_r S_c$ line (limits of $N_b$ when $E_a \to \infty$ ). As the binding strength increases, the $N_b$ allocation function approaches both limits. However, a smooth transition from the ligand-dominated to the receptor-dominated type of equilibrium takes place. Hence, for strong binding (e.g. $E_a > 5$ in the case of parameters chosen in Fig. 5), the $N_b$ allocation function is virtually cornered between the two limits, with a crossover at $S_c^*$.

While $S_c^* < 0.5 S_t$, the $N_b$ functions are linear for small binding strengths ( $E_a \ll 1$ ). Here, $N_b^{LIN}$ matches the real solution very well. For intermediate binding strengths, $N_b^{LIN}$ is a good approximation only at small sizes of the contact zone. For large binding strengths $N_b^{LIN} \to \rho_r S_c$, so it again becomes representative of the real $N_b$ (until the $N_t$ line is intercepted at $S_c^*$), as can be seen in the left panel of Fig. 5.

*(b) $S_c^* > 0.5 S_t$ regime.* In this regime the system is always in the receptor-dominated equilibrium (right panel in Fig. 5). Thus the region where solutions for $N_b$ can be found is limited only by the $\rho_r S_c$ line. Furthermore, $N_b$ is almost linear for any choice of parameters. Increasing the binding strength causes the saturation of $N_b$ to the $\rho_r S_c$ line in the whole range of available sizes of the contact zone (*e.g.* curves with $E_a > 5$ cannot be distinguished from the $\rho_r S_c$ boundary).

Regardless of the binding strength, when $S_c^* > 0.5 S_t$, the maximum deviation of $N_b^{LIN}$ from $N_b$ is found to be less than 5% and arises for binding strengths $E_a \cong 1$ at small sizes of the contact zone. It is in this limit that the approximation given by $N_b^0$ can be used as a substitute.

### *2. Dependence of the allocation functions on the binding strength of the ligand-receptor pair*

In this presentation, the concentrations of the ligands and the receptors are preset, as is $S_c^*$. Furthermore, the size of the contact zone must be chosen. The relation between these system parameters, as shown in previous sections, determines whether the system will be in a ligand-dominated or a receptor-dominated equilibrium. Thus when $N_b$ is plotted as a function of the ligand-receptor binding strength, the entire resulting curve is a set of only one type of equilibria (Fig. 6). In this case, for small to intermediate values of the binding strength, the allocation function for bound ligands experiences rapid almost linear growth that continuously deviates into a saturation regime defined by one of the two limits at $E_a \to \infty$. In principle, the linear regime can be characterized by expanding around the inflection point of the allocation functions that are almost symmetrical sigmoids. Due to the complexity of such expressions, a simple but successful procedure has been undertaken for determining the alternative expansion point, the coordinates of which will be denoted as $N_b^M$ and $E_a^M$. The ordinate $N_b^M$ is determined as the value intermediate between the limits of the allocation function at $-\infty$ and $+\infty$, with $N_b \to 0$, when $E_a \to -\infty$. The abscise $E_a^M$ is the binding strength at which the $N_b$ function intersects the $N_b^M$ value (see Fig. 6). However, for both types of equilibria, it is possible to approximate the linear regime of the allocation function by:

$$N_b \cong N_b^{SIG} = N_b^M \left[ k \cdot \left( E_a - E_a^M \right) + 1 \right] \ . \tag{A3}$$

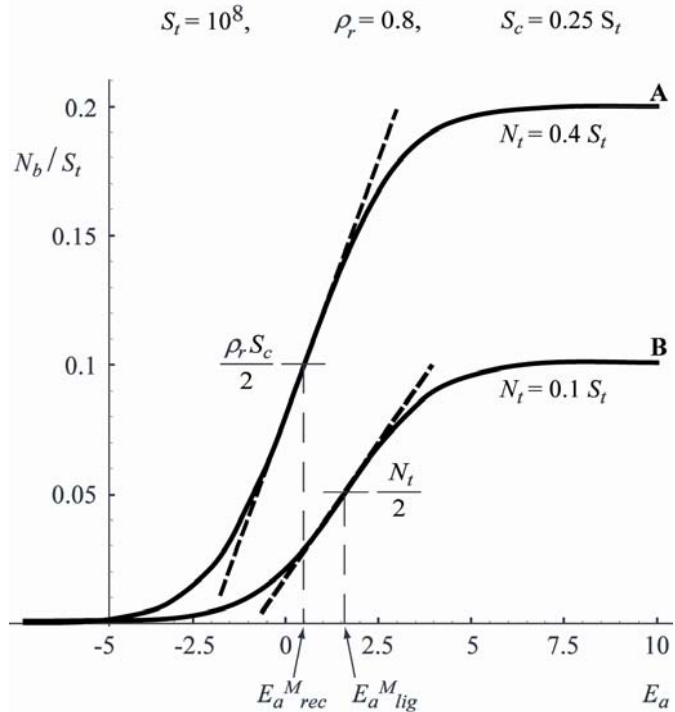

**FIG. 6:** The allocation function of bound ligands as a function of the ligand-receptor binding strength. The real solution $N_b$ and the approximate $N_b^{SIG}$ are presented with full and short-dashed lines, respectively. Points used for the expansion can be associated with the intersection of $N_b$ curves with the long dashed lines. The curves denoted by A belong to the case $S_c < S_c^*$ and are a representative of a set of receptor-dominated equilibria. Curves denoted by B belong to the case when $S_c > S_c^*$ and are a construction of ligand-dominated equilibria.

Due to different limits at $+\infty$, the values of $N_b^M$, $E_a^M$ and $k$ will depend on whether the system is in the ligand or in the receptor dominated equilibria.

*(a) Receptor-dominated equilibria.* The coordinates of the expansion point are found to be:

$$N_b^M = \frac{\rho_r S_c}{2}, \qquad E_a^M = \ln \frac{2S_t - 2N_t - \rho_r S_c}{2N_t - \rho_r S_c}, \tag{A4a}$$

$$k = \frac{(2N_t - \rho_r S_c)(2N_t + \rho_r S_c - 2S_t)}{8N_t^2 - 8N_t S_t + 2\rho_r S_c S_t}. \tag{A4b}$$

For curves representing receptor-dominated equilibria (curve A in Fig. 6), increasing the number of ligands in the vesicle ($N_t$) while keeping the coverage density ($\rho_r$) and the size of the contact zone ($S_c$) constant, results in a shift of $E_a^M$ to smaller energies. As $N_b$ must converge to a constantly maintained value of $\rho_r S_c$, and the slope of $N_b^{SIG}$ is not significantly altered, $N_b$ also converges to its limiting value at smaller values of $E_a$. On the other hand, maintaining $N_t$ constant while increasing either $\rho_r$ or $S_c$ leads to a convergence of $N_b$ at increased values, which considerably increases the slope $k$. Nevertheless, the saturation of $N_b$ is reached more slowly in both cases. Interestingly, an increase of $\rho_r$ (at constant $N_t$ and $S_c$) results in a shift of $E_a^M$ to smaller values, whereas an increase of $S_c$ (at constant $N_t$ and $\rho_r$) has the opposite effect.

*(b) Ligand-dominated equilibria.* The coordinates of the expansion point are found to be:

$$N_b^M = \frac{N_t}{2}, \qquad E_a^M = \ln \frac{2S_t - N_t - 2\rho_r S_c}{N_t - 2\rho_r S_c}, \tag{A5a}$$

$$k = \frac{(N_t - 2\rho_r S_c)(N_t + 2\rho_r S_c - 2S_t)}{8\rho_r^2 S_c^2 - 8\rho_r S_c S_t + 2N_t S_t}. \tag{A5b}$$

For this type of equilibrium (curve B in Fig. 6) adding receptors to the surface or increasing the size of the contact zone, while maintaining $N_t$ constant, results in convergence of $N_b$ to $N_t$ at smaller values of $E_a$. Although the slope of $N_b^{SIG}$ remains almost unaltered, raising either $\rho_r$ or $S_c$ results in a shift of $E_a^M$ to smaller values. Preparing vesicles with higher ligand concentration while keeping the size of the contact zone (*e.g.* the reduced volume of the vesicle) or the substrate composition constant, will increase the saturation level of $N_b$. In this case, the convergence is achieved at higher values of both $E_a$ and $E_a^M$. Nevertheless, the slope $k$ is increased with a higher content of ligands.

*(c) Relevance to applications.* Together with the convergence limits, which are usually reached at ligand-receptor binding strengths of the order of $10\,k_B T$, the developed approximations cover most of the parameter space in which $N_b$ should be determined. As, can be seen from Table II, expansions of the type $N_b^{SIG}$ are good approximations to $N_b$ and can be used when the somewhat simpler $N_b^{LIN}$ is inappropriate. This is particularly important for the intermediate range of the ligand-receptor binding strengths. In addition, for low coverage or small contact zones, the slope coefficient $k$ in $N_b^{SIG}$ can be considerably simplified.

**Table II.** Regions of parameters for the applicability of different approximate relations of the allocation function of the bound ligands. The critical size of the contact zone is a ratio between the total number of ligands in the vesicle and the density of receptors on the substrate ($S_c^* = N_t / \rho_r$). $S_c < S_c^*$ is indicative of a receptor-dominated type of equilibrium whereas $S_c > S_c^*$ results in a ligand-dominated equilibrium. $N_b^{LIN}$ is defined in text by Eq. (A1). $N_b^{SIG}{}_{rec}$ and $N_b^{SIG}{}_{lig}$ are defined by the Eq. (A3). The subscripts *lig* or *rec* signify the use of parameters given in Eq.(A4) and Eq. (A5), respectively.

| $E_a$ | $S_c^* > 0.5 S_t$ | | $S_c^* < 0.5 S_t$ | |
|---|---|---|---|---|
| | $S_c < S_c^*$ | $S_c < S_c^*$ | | $S_c > S_c^*$ |
| very low | $N_b^{LIN} \to \rho_r S_c N_t / S_t$ | | | |
| Low-medium | $N_b^{LIN}$ | | $N_b^{SIG}{}_{rec}$ | $N_b^{SIG}{}_{lig}$ |
| strong | $N_b^{LIN} \to \rho_r S_c$ | | | $N_t$ |

The allocation function for free ligands in the contact zone ($N_f$) and the allocation function for the total number of free ligands in the vesicle ($N_{free}$) share the same $E_a^M$. Hence, their expansions can be calculated by the use of the density equation (3), in which $N_b$ should be replaced by the appropriate $N_b^{SIG}$.